# Data-Driven Optimal Sensor Placement for High-Dimensional System Using Annealing Machine

*Tomoki Inoue* [1], *Tsubasa Ikami* [2,3], *Yasuhiro Egami* [4], *Hiroki Nagai* [3], *Yasuo Naganuma* [5], *Koichi Kimura* [5], *Yu Matsuda* [1,6,*]

1. Department of Modern Mechanical Engineering, Waseda University, 3-4-1 Ookubo, Shinjuku-ku, Tokyo, 169-8555, Japan

2. Department of Aerospace Engineering, Tohoku University, 6-6-01 Aoba, Arakaki, Aoba-ku, Sendai, Miyagi-Prefecture 980-8579, Japan

3. Institute of Fluid Science, Tohoku University, 2-1-1 Katahira, Aoba-ku, Sendai, Miyagi-prefecture 980-8577, Japan

4. Department of Mechanical Engineering, Aichi Institute of Technology, 1247 Yachigusa, Yakusa-Cho, Toyota, Aichi-prefecture 470-0392, Japan

5. Quantum Software Project, Quantum Laboratory, Research Unit of Fujitsu Research, Fujitsu Ltd. Kanagawa 211-8588, Japan

6. Japan Science and Technology Agency, PRESTO, Saitama 332-0012, Japan

* Corresponding author: y.matsuda@waseda.jp

**Abstract**

We propose a novel method for solving optimal sensor placement problem for high-dimensional system using an annealing machine. The sensor points are calculated as a maximum clique problem of the graph, the edge weight of which is determined by the proper orthogonal decomposition mode obtained from data based on the fact that a high-dimensional system usually has a low-dimensional representation. Since the maximum clique problem is equivalent to the independent set problem of the complement graph, the independent set problem is solved using Fujitsu Digital Annealer. In contrast to the greedy method, which selects the optimal point at each step and never reconsider the point selected previously, the proposed method is superior because it is able to find the optimal set of points. As a demonstration of high dimensional system, the pressure distribution measured by the pressure-sensitive paint method, which is an optical flow diagnose method, is reconstructed from the pressure data at the calculated sensor points. The root mean square errors (RMSEs) between the pressures measured by pressure transducers and the pressures reconstructed from the proposed method, an existing greedy method, and random selection method are compared. The similar RMSE is achieved by the proposed method using approximately 1/5 number of sensor points calculated by the existing method. This method is of great importance as a novel approach for optimal sensor placement problem and a new engineering application of an annealing machine.

1. Introduction

Optimal sensor placement problem has received attention for a long time, as a fast data acquisition, analysis, and decision in high performance control for complex system can be archived with a small number of measurements at optimal places [1, 2] and there are many possible applications such as thermo-fluid sensing [3-5], room environment monitoring [6], structural health monitoring [7-10], robotic sensing [11], chemical plant sensing [12], agricultural environment monitoring [13], and so on. Moreover, the reduction in the number of sensors generally leads to reduction in the cost of measurement systems. For a moderately sized system, optimal sensor placement is estimated by information theoretic frameworks such as convex optimization [2, 14], Bayesian approaches [6, 15], and deep learning frameworks [16], and so on. However, these methods have a high computational and are inappropriate for applying to high-dimensional systems. Though the computationally efficient convex optimization algorithm is proposed [10], the convexity is lost when measurement noise is taken into account [17]. From the view point of signal reconstruction, greedy optimizations based on QR factorization [3] and the determinant of matrix [5, 18, 19] are proposed as scalable methods. However, it is well known that greedy optimization fails to find the globally optimal solution. In these methods [3, 18], the demonstration is conducted using time-series data of $360 \times 180$ spatial grid and 520 or 832 snapshots. Dramatic advances in data acquisition technology have increased the amount of data by a factor of 2 to 100 [20-22]. A new approach is required to find the globally optimal solution for these high-dimensional data containing a large amount of measurement noise. Since optimal sensor placement problem is one of a combinatorial problem that determine choices of $q$ sensor points out of $n$-dimensional system, we focus on an annealing machine, specifically Fujitsu Digital Annealer (DA) [23, 24], to solve the problem. DA is designed as a dedicated architecture for solving fully connected quadratic unconstrained binary optimization (QUBO) problems. DA has been applied to solve combinatorial optimization problems such as materials science [25, 26], life science [27], and so on. Details of the operating principle and the performances of DA are discussed in references [23, 24].

In the present paper, we propose a novel approach to solve optimal sensor placement problem for high-dimensional system using DA, because the problem can be mapped to an Ising model which is equivalently a QUBO problem [28, 29]. We use second-generation DA (DA2), which is capable of solving up to 8192 bits of problem space. As a demonstration, the proposed method is applied to the high-dimensional fluid dynamic data of $780 \times 780$ spatial grid and

8192 snapshots, which is approximately 100 times larger than the data used in the existing studies [3, 18]. We use the pressure distribution behind the square cylinder data measured in our previous study [30]. The noise in the data was suppressed based on the data at the optimally placed points calculated by the greedy optimization method [18] in the study. That is, the denoised pressure distribution was reconstructed by the superposition of multiple modes to match the data at the optimally placed points. In this study, the reconstructed pressure data is compared with the pressure data measured by pressure transducers. The estimation errors of the reconstructed data by the proposed method and the existing method are compared.

## 2. Proposed Method for Solving Optimal Sensor Placement Problem

The high-dimensional system can be expressed by small number of measurements at optimally placed points due to the fact that high-dimensional system often has a low-dimensional representation based on tailored/data-driven bases such as proper orthogonal decomposition (POD) [31, 32] and dynamic mode decomposition (DMD) [33, 34]. In this study, we used snapshot POD [31] as a data-driven base following the Refs. [3] and [18] to compare the result obtained by the proposed method with that by the greedy method [18]. We consider $n$-dimensional system with $m$ measurements, that is, we consider an observation data matrix $\mathbf{X} \in \mathbb{R}^{n \times m}$ in this study. The singular value decomposition (SVD) of the observation data matrix $\mathbf{X}$ provides the following matrix decomposition:

$$\mathbf{X} = \mathbf{U}\mathbf{\Sigma}\mathbf{V}^\top, \tag{1}$$

where $\mathbf{U} \in \mathbb{R}^{n \times n}$ and $\mathbf{V} \in \mathbb{R}^{m \times m}$ are unitary matrices and the superscript ⊤ denotes the transpose. The matrix $\mathbf{\Sigma} \in \mathbb{R}^{n \times m}$ is a diagonal matrix, each element of which is a singular value arranged in descending order. The matrix $\mathbf{U}$ represents the POD modes as

$$\mathbf{U} = [\mathbf{\psi}_1, \mathbf{\psi}_2, \cdots, \mathbf{\psi}_m]. \tag{2}$$

where $\mathbf{\psi}_i$ indicates $i$-th POD mode. In other words, the $i$-th column of $\mathbf{U}$ corresponds to the $i$-th POD mode $\mathbf{\psi}_i$. The data can be approximately represented by a rank-$r$ truncated SVD ($r < m$) as

$$\mathbf{X} \approx \widetilde{\mathbf{U}}\widetilde{\mathbf{\Sigma}}\widetilde{\mathbf{V}}^\top, \tag{3}$$

where $\tilde{\mathbf{U}} \in \mathbb{R}^{n \times r}$, $\tilde{\mathbf{\Sigma}} \in \mathbb{R}^{r \times r}$ and $\tilde{\mathbf{V}} \in \mathbb{R}^{m \times r}$ are truncated matrices and $\tilde{\mathbf{U}}$ represents the truncated POD modes as $\tilde{\mathbf{U}} = [\boldsymbol{\psi}_1, \boldsymbol{\psi}_2, \cdots, \boldsymbol{\psi}_r]$.

The observation vector $\mathbf{y} \in \mathbb{R}^q$ is approximately represented by the data-driven base, POD modes, and their coefficient vector $\mathbf{a} \in \mathbb{R}^r$ as follows [3]:

$$\begin{aligned}\mathbf{y} &\approx \mathbf{C}\tilde{\mathbf{U}}\mathbf{a} \\ &= \boldsymbol{\Theta}\mathbf{a}\end{aligned}, \quad (4)$$

where $\mathbf{C} \in \mathbb{R}^{q \times n}$ represents the optimal sensor position, each element of which is 0 or 1, and each row vector of $\mathbf{C}$ has a single unity element that indicates a sensor position. Here, $q$ is the number of sensors, and $r$ is the truncation rank of the POD mode. The problem is to find a sparse representation of sensor position matrix $\mathbf{C}$ satisfying Eq. 4. The sensor matrix $\mathbf{C}$ selects points from the spatial data $\tilde{\mathbf{U}}\mathbf{a}$, but focusing on $\boldsymbol{\Theta}(= \mathbf{C}\tilde{\mathbf{U}})$, it can also be interpreted as selecting row vectors $\mathbf{u}_j$ from $\tilde{\mathbf{U}}$. Each element of the vector $\mathbf{u}_j \in \mathbb{R}^r$ corresponds to the element of each POD mode at the selected point $j$. The selected points represent the spatial signal well and are suitable to be employed as sensor point. From the view point of selecting a small number of sensor points, however, the vectors with high similarity are unnecessary because such points indicate similar information. It is also noted that the contribution of each POD mode to the representation of the phenomenon is different; then, the row vectors $\hat{\mathbf{u}}_j \in \mathbb{R}^r$ of $\tilde{\mathbf{U}}\tilde{\mathbf{\Sigma}}$ are considered, where each POD mode is weighted by the singular value $\tilde{\mathbf{\Sigma}}$. Then, we consider an edge-weighted undirected graph $\mathcal{G}$, each vertex (node) of which corresponds to each point (each row vector $\hat{\mathbf{u}}_j$), and all vertices are connected each other by edges. Though there are several choices of the weight of the edge, we consider the $\ell_2$ norm of the cross product of two weighted row vectors $\hat{\mathbf{u}}_k$ and $\hat{\mathbf{u}}_l$ corresponding to the endpoints of the edge as the weight $w(\hat{\mathbf{u}}_k, \hat{\mathbf{u}}_l)$. Then, the weight is described as follows:

$$w(\hat{\mathbf{u}}_k, \hat{\mathbf{u}}_l) = \|\hat{\mathbf{u}}_k \times \hat{\mathbf{u}}_l\|_2, \quad (5)$$

where $\|\cdot\|_2$ denotes the $\ell_2$ norm. The subscripts $k$ and $l$ ($k \neq l$) are integers representing $k$-th and $l$-th rows of $\tilde{\mathbf{U}}\tilde{\mathbf{\Sigma}}$, respectively. The weight will be larger, when the similarity between $\hat{\mathbf{u}}_k$ and $\hat{\mathbf{u}}_l$ is lower and the $\ell_2$ norm of each vector is larger. Then, we find a set of vertices connected by edges with a large weight $w$ value to find optimal sensor points. We select $q$ optimal sensor points as a maximum clique problem from the undirected graph $\mathcal{G}_c$, each weight of which is larger than $c$. It is noted that the threshold for the weight $c$ determine the number of sensor points $q$; thus, the smaller $c$ is, the bigger the size of maximum clique and vice versa.

The maximum clique problem is equivalent to the independent set problem of the complement graph $\overline{\mathcal{G}_c}$ of the graph $\mathcal{G}_c$ and is known as a NP-hard problem [28, 29]. In this study, we focus on the independent set problem of the complement graph $\overline{\mathcal{G}_c}$ and solve it using DA2. The Ising Hamiltonian $\mathcal{H}$ that solves it is expressed as follows:

$$\mathcal{H} = -\Lambda_1 \sum_{\zeta \in \overline{\mathcal{G}_c}} x_\zeta + \Lambda_2 \sum_{(\zeta,\xi) \in \Omega} x_\zeta x_\xi , \tag{6}$$

where $\Omega$ represents the edge set of $\overline{\mathcal{G}_c}$ and $x$ is a binary bit variable. The subscript $\zeta$ of $x_\zeta$ indicates $\zeta$-th vertex. We set $x_\zeta = 1$ and $x_\zeta = 0$ for the vertex belonging to and not to $\overline{\mathcal{G}_c}$, respectively. The constants $\Lambda_1$ and $\Lambda_2$ determine the weight of each term. This equation is equivalently a QUBO problem. The first term of $\mathcal{H}$ represents the number of vertices belonging to $\overline{\mathcal{G}_c}$. On the other hand, the second term represents the penalty for selecting vertices both end of an edge. By minimizing $\mathcal{H}$ under the condition of $\Lambda_1 < \Lambda_2$, $\Lambda_1 = 1$ and $\Lambda_2 = 2$ in this study, we can obtain independent set of the complement graph $\overline{\mathcal{G}_c}$, and then obtain $q$ optimal sensor points as a maximum clique of the graph $\mathcal{G}_c$. It is expected that the proposed method enables us to extract more appropriate sensor points than the existing greedy method, because the proposed method finds optimal set of sensor points whereas the greedy method selects sensor point depending only on the selections made previously, but not on future selection (the greedy method does not reconsider the points selected previously) as shown in **Fig. 1**.

3. **Validation of Proposed Method**

As a validation of the proposed method, we applied the proposed method to the NOAA OISST V2 global ocean surface temperature data, which are provided by the NOAA PSL [35]. The data consists of $360 \times 180$ spatial grid and 1532 snapshots (weekly data from Jan. 1990). This data set is relatively small and is frequently used in the existing studies [3, 18], in which the temperature distribution is reconstructed from the POD modes and the selected sensor points. It is reported that the reconstruction error by the greedy method is equivalent to that by the convex approximation [18]. Then, we have compared the reconstruction error by the proposed method with that by the greedy method [18]. Following Ref. [18], the sensor points were calculated based on the POD modes with the truncation rank of $r = 10$. The coefficient vector $\mathbf{a}_l$ in Eq. (4) at the time step of $l$ is calculated as

$$\mathbf{a}_l = \mathbf{\Theta}^\dagger \mathbf{y}_l , \tag{7}$$

where $\boldsymbol{\Theta}^{\dagger}$ is pseudo-inverse of $\boldsymbol{\Theta}$. The reconstructed data $\tilde{\mathbf{x}}_l$ at the time step of $l$ is represented as

$$\tilde{\mathbf{x}}_l = \tilde{\mathbf{U}} \mathbf{a}_l . \tag{8}$$

The reconstructed error $e_{\text{reconst}}$ is defined as

$$e_{\text{reconst}} = \frac{1}{N} \sum_{l=1}^{N} \frac{\|\mathbf{x}_l - \tilde{\mathbf{x}}_l\|_2^2}{\|\mathbf{x}_l\|_2^2} , \tag{9}$$

where $\mathbf{x}_l$ is the original data at the time step of $l$ and $N$ is the total number of the snapshots ($N = 1532$ in this study).

**Figure 2** shows the typical results of the selected sensor points and the reconstructed temperature distribution of third week of April 2000 for the number of sensor points of $q = 66$. The reconstruction errors are shown in **Fig. 3**. Since the problem space that can be treated by DA2 is limited, we selected sensor points of $q = 9$ to 192 from uniformly distributed 2000 points. The data obtained by the greedy method and a random selection method, in which the sensor points were randomly selected, are also shown for comparison. Since the results by the random selection method strongly depends on the random seed used, the mean values and the standard deviations calculated from 32 independent trials are shown in Fig. 3. As shown in Fig. 2, the temperature distributions were successfully reconstructed by these methods. The reconstruction error by the proposed method smaller than that by the random selection method in this study. The reconstructed error by the greedy method is smaller than those by the random selection method and the proposed method when the number of sensor points $q < 45$. This is due to the limitation of the problem space of DA2. Interestingly, the errors by the random selection method and the proposed method become smaller that by the greedy method for the number of sensor points $q > 45$. This is because the many sensor points that are spatially close to each other are selected in the greedy method as shown in Fig. 2(c) and these points would not improve the reconstruction performance.

## 4. Application to High Dimensional Problem
### 4.1. Experimental methods

Next, we consider high-dimensional data containing a large amount of measurement noise. The pressure distributions used in this study were measured by the pressure-sensitive paint (PSP) method [36, 37] in the previous study [30, 38]. The schematic illustration of the measurement is shown in **Fig. 4**. The PSP coating was applied to a turn table on which a square cylinder was placed. The Kármán vortex street behind the square cylinder (width: 40 mm, length: 40 mm, and height: 100 mm) was measured by a high-speed camera. The exposure time was set to 1/3000 s and the frame rate of the camera was 1000 fps. A mean velocity of flow was 50 m/s; thus, the Reynolds number based on the width of the square cylinder was $1.1 \times 10^5$.

In the PSP method, a PSP coating containing pressure-sensitive phosphorescent dyes is applied to a surface of interest. When the PSP coating is illuminated with a light of proper wavelength, it emits a luminescence, the intensity of which varies depending on local pressure. Then, pressure distribution can be measured by detecting the variation of the PSP emission with a camera. The relation between pressure and the luminescent intensity is expressed by the Stern-Volmer relation [37, 39],

$$\frac{I_{\text{ref}}}{I} = A + B \frac{p}{p_{\text{ref}}} \tag{10}$$

where $I$ and $p$ are respectively the luminescent intensity and pressure. The subscript, ref, indicates a reference condition which is an atmospheric condition in this study; thus, the $I_{\text{ref}}$ is the luminescent intensity of the PSP at an atmospheric condition $p_{\text{ref}}$. The constants $A$ and $B$ are called as the Stern-Volmer constants satisfying the constraint condition of $A + B = 1$ [40, 41] and are determined by a calibration test. In the flow measurement, the "wind-on" images $I$ were captured at a mean velocity of 50 m/s and "wind-off" images $I_{\text{ref}}$ were immediately captured after the wind tunnel was stopped. The measured data size was $780 \times 780$ spatial grid and 8192 snapshots in this study.

The pressures ($P_1$, $P_2$, $P_3$, and $P_4$) downstream of the square cylinder were measured by pressure transducers through pressure taps (diameter: 0.5 mm) at four different places. The measured pressures by the pressure transducers were compared with the PSP results. More details are written in our previous studies [30, 38].

### 4.2. Signal reconstruction method

PSP measurement with small pressure fluctuations of less than 0.1 kPa near an atmospheric pressure is a central challenge because the variation of the luminescent intensity of ~0.1% has to be detected, and this small intensity change is usually buried in camera noise. Thus, we proposed a signal reconstruction method to extract the pressure distribution from the noticeable noise in the previous study [30]. In this paper, we explain the brief procedure of the reconstruction method. First, the POD mode is calculated from the time-series PSP measurement data. The obtained POD modes are used as modal basis. Second, optimal sensor placement problem is calculated based on the POD mode by the greedy method [18] or the proposed method. Third, the time-series pressure data at optimally placed points is filtered to remove the noise. Finally, the coefficient vector of the POD modes is determined by minimizing the difference between the above-mentioned filtered pressure data and the reconstructed pressure at the optimal points under the assumption that the coefficient vector is sparse. Here, the reconstructed pressure distribution $\tilde{\mathbf{p}}_l$ at time $\tau = l\Delta t$, where $l$ is an integer and $\Delta t$ is a time interval of the measurement, is expressed as

$$\tilde{\mathbf{p}}_l = \tilde{\mathbf{U}}\boldsymbol{\alpha}_l = \sum_{i=1}^{r} \alpha_{l,i}\boldsymbol{\psi}_i \; , \tag{11}$$

where $\boldsymbol{\alpha}_l = (\alpha_{l,1}, \alpha_{l,2}, \cdots, \alpha_{l,r})^\top$ and the amplitude $\alpha_{l,i}$ for the $i$-th POD mode $\boldsymbol{\psi}_i$ at the time $l\Delta t$. It is noted that the pressure distribution $\tilde{\mathbf{p}}_l$ is represented as a column vector, the size of which is $n = n_\mathrm{h} \times n_\mathrm{v}$ ($n_\mathrm{h}$ and $n_\mathrm{v}$ are the number of horizontal and vertical pixels of image, respectively). That is, the two-dimensional pressure distribution matrix is reshaped into a column vector. We consider the time series pressure at the optimal sensor points. Since each pressure data contains a large amount of noise due to unsteady PSP measurement with a short camera exposure and small pressure fluctuation, a spatial mean filter around the optimal point is used as a simple filter in this study. The filtered pressure data at the optimally placed points and time $\tau = l\Delta t$ is denoted as $\boldsymbol{\phi}_l$. The amplitude $\boldsymbol{\alpha}_l$ in Eq. 11 is determined by minimizing the difference between the filtered pressure data $\boldsymbol{\phi}_l$ and the reconstructed pressure $\mathbf{C}\tilde{\mathbf{p}}_l$ at the optimally placed points. The matrix $\mathbf{C}$ represents the optimal sensor position. The amplitude $\boldsymbol{\alpha}_l$ is determined by the LASSO (least absolute shrinkage and selection operator) [42-44] as follows:

$$\boldsymbol{\alpha}_l^* = \operatorname*{argmin}_{\boldsymbol{\alpha}_l} \frac{1}{2q} \|\boldsymbol{\phi}_l - \mathbf{C}\tilde{\mathbf{U}}\boldsymbol{\alpha}_l\|_2^2 + \lambda\|\boldsymbol{\alpha}_l\|_1 \tag{12}$$

where $\boldsymbol{\alpha}_l^*$ is estimated coefficient vector and $\|\cdot\|_1$ denotes the $\ell_1$ norm. The parameter $\lambda$ weights the importance of sparsity and is determined by the 10-fold cross-validation using the one-standard-error rule [43-45].

### 4.3. Results and Discussion

As a demonstration, we have applied the proposed method to the reconstruction of pressure distribution induced by the Kármán vortex street behind a square cylinder. The pressure distribution on the turn table was measured as shown in **Fig. 5**, where the pressure $p$ is normalized by an atmospheric pressure $p_{\text{ref}}$. More detail data is shown in Fig. S1. As shown in the figure, the measured pressure distributions contain a noticeable noise due to the small pressure variation of approximately 1%. Then, the noiseless distributions were reconstructed based on the data at optimal points by the method described in Sec. 4.2. In this study, we have compared the results obtained by the proposed method with those by the random selection method. We also compared the results with those by the existing greedy method [18], because the accuracy archived by the greedy method is similar to other existing methods and the computational cost is smaller than the other methods as shown in Ref. [18]. As discussed in Eq. 3, the truncated POD modes were considered and the truncation rank was calculated as 100 following Gavish and Donoho [46]. However, the truncation rank of $r = 128$ was used for redundancy to calculate optimal sensor placement problem. Since the problem space that can be treated by DA2 is limited, we selected sensor points of $q = 30, 60, 96, 128, 215,$ and $276$ from uniformly distributed 2000 points. For all cases, we obtained the solution with $\mathcal{H} = -q$, which did not contain the penalty term expressed in Eq. 6; thus, the choice of the constants $\Lambda_1$ and $\Lambda_2$ did not affect the results. The processing time containing input/output operations for the calculation in DA2 were 10 s. **Figure 6** shows the typical results of sensor points for $q = 60$ calculated by the random selection, the existing method, and the proposed method. Figure 6(b) shows that 3/4 of sensors were placed around the boundary of the PSP (turn table) and 1/4 of them were placed on the downstream of the square cylinder by the existing method. On the other hand, most of sensors were placed on the downstream of the cylinder by the proposed method as shown in Fig. 6(c). The downstream of the cylinder is important to characterize the distribution because the vortices were passing through this area. Similar trends were also observed in other conditions of the calculations (see Fig. S2). Though sensor points close to each other were selected by the existing greedy method (particularly around the boundary), the selected points are a certain distance apart from each other in the proposed method. Measurements at the points close to each other are expected to provide similar results and are not considered informative. **Figure 7** shows the reconstructed pressure distributions based on the sensor points calculated by the random selection method, the existing method, and the proposed method. The pressure is normalized by an atmospheric pressure $p_{\text{ref}}$. The reconstructed data by other conditions are shown in Figs. S3 – S5. As shown in Fig. 7, the noise was significantly reduced in all methods. However, the pressures were clearly different each other at the areas indicated by dashed circles. It is difficult to determine which method is more appropriate from the distribution. Then, we compared the root mean square error (RMSE) between the pressure measured by the pressure transducer and the reconstructed pressure at the pressure tap. The results for 400 time step data are shown in **Fig. 8**. The RMSE of the raw data was 0.045, which is not shown in Fig. 8. In this study, we reconstructed noiseless pressure distribution from noisy data; thus, the result obtained by the spatial averaging filter of $3 \times 3$ pixels, which is a common noise reduction method, is also shown for comparison. For the random selection method, the mean values and the standard deviations calculated from 32 independent trials are shown in the figure. The RMSEs obtained by the random selection, existing greedy, and the proposed methods were dramatically reduced from those by the raw data and the averaging filter. This result indicates the usefulness of our noise reduction algorithm. Now, we focus on the RMSEs obtained by the random selection, existing greedy, and the proposed methods. The RMSEs decreased with the increase of the number of sensor points, and the decrease of the RSMEs became small for the number of sensor points $q > 128$, which is the number of the POD mode used for the calculation. The RMSE of the proposed method was smaller than those of the existing method and of the random selection method. Moreover, the RMSE of the proposed method with $q = 60$ was similar to that of the existing method with $q = 276$. This means that the reconstruction with similar accuracy was archived by approximately 1/5 number of sensor points; in other words, the proposed method extracted the points efficiently representing the distribution compared with other methods. The proposed method is powerful method to solve optimal sensor placement problem and will contribute a fast data acquisition, analysis, and decision in high performance control for complex system with a small number of measurements at optimal places.

## 5. Conclusion

Sensor placement problem has received much attention due to attractive engineering applications such as observation, estimation and prediction of high-dimensional system. In this study, we propose a novel method for solving optimal sensor placement problem for high-dimensional system using an annealing machine. The sensor points are calculated as a maximum clique problem of the graph, the edge weight of which is determined by the proper orthogonal decomposition (POD) mode obtained from data, because a high-dimensional system usually has a low-dimensional representation. Since the maximum clique problem is equivalent to the independent set problem of the complement graph, the independent set problem is solved using second-generation Fujitsu Digital Annealer (DA2). As a validation of the proposed method, we considered the NOAA-SST sensor problem. The temperature distribution was successfully reconstructed based on the sensor points estimated by the proposed method. The reconstruction error by the proposed method was smaller than that by the existing greedy method when the number of sensor points was large. As a demonstration of the proposed method, the pressure distribution induced by the Kármán vortex street behind a square cylinder is reconstructed based on the pressure data at the calculated sensor points. Though about 60% of the sensor points calculated by the existing greedy method concentrated on the boundary of the PSP, most of those calculated by the proposed method were placed on the behind of the square cylinder. We compared the root mean square errors (RMSEs) between the pressure measured by pressure transducers and the reconstructed pressures (calculated from the proposed method and an existing greedy method) at the same place. As the result, the similar RMSE is achieved by the proposed method using approximately 1/5 number of sensor points obtained by the existing method. The sensor points extracted by the proposed method represent the pressure distribution more efficiently than the points extracted by other methods. We believe that the method is of great importance as a novel approach for optimal sensor placement problem and a new engineering application of an annealing machine.

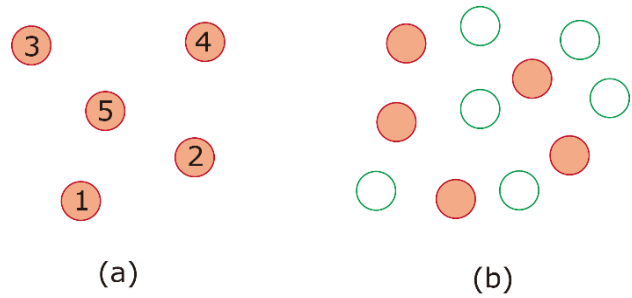

**Figure 1.** Schematic illustrations of sensor point selection by (a) greedy and (b) proposed method. The proposed method extracts the set of sensor points. On the other hand, greedy method selects a point one-by-one and never reconsider the previous selection.

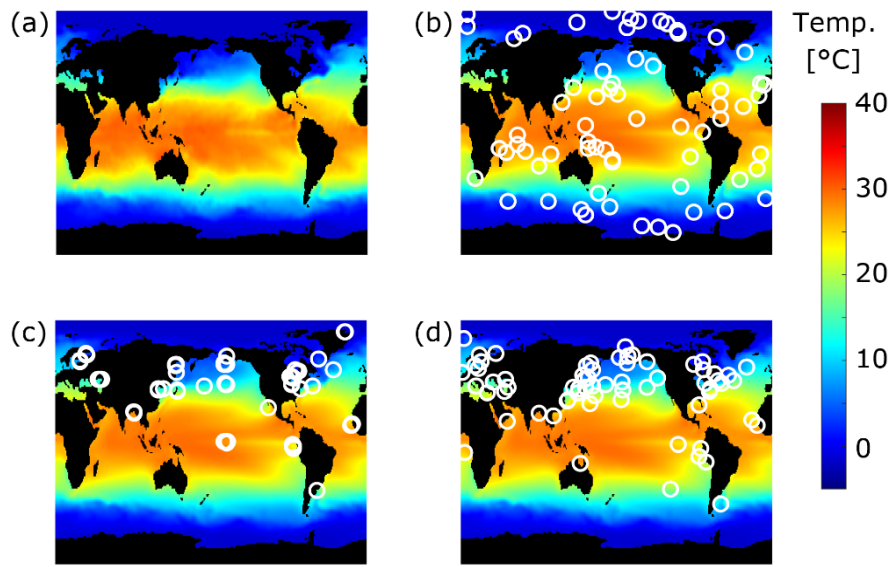

**Figure 2.** Typical sensor positions and reconstructed temperature distributions of third week of April 2000 for the number of sensor points of $q = 66$ estimated by (b) the random selection method, (c) the existing greedy method, and (d) the proposed method. (a) the original temperature distribution. The sensor position is shown in white circle.

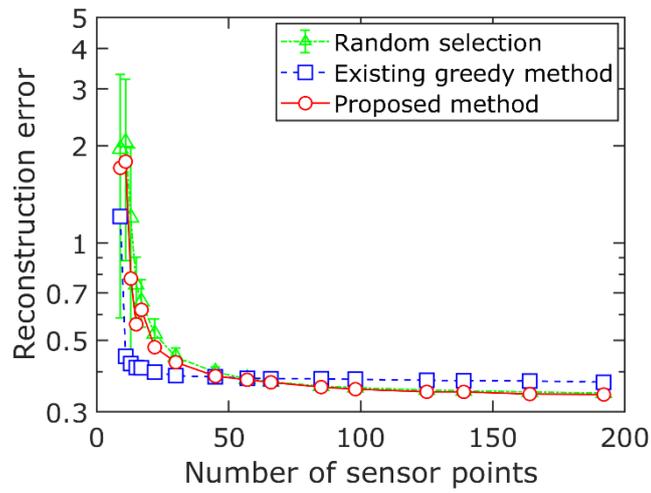

**Figure 3.** The comparison of reconstruction error between the proposed method and other methods

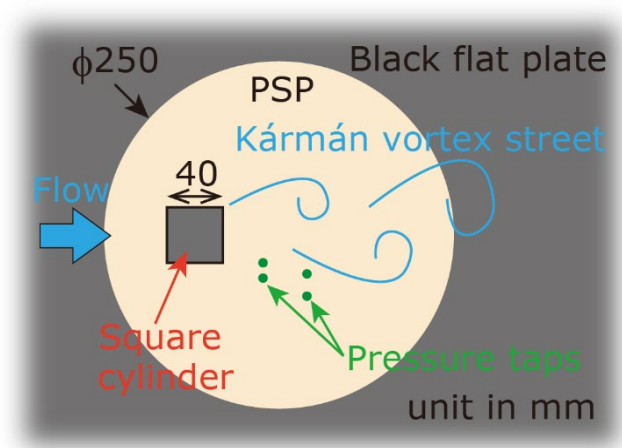

**Figure 4.** Schematic illustration of PSP measurement in this study

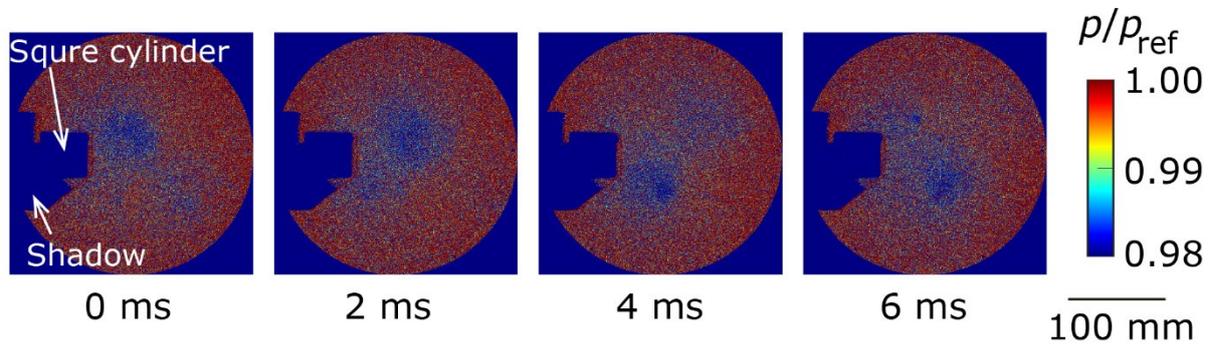

**Figure 5.** Typical pressure distributions (raw data). The pressure $p$ is normalized by an atmospheric pressure $p_{\text{ref}}$. The flow direction is left to right.

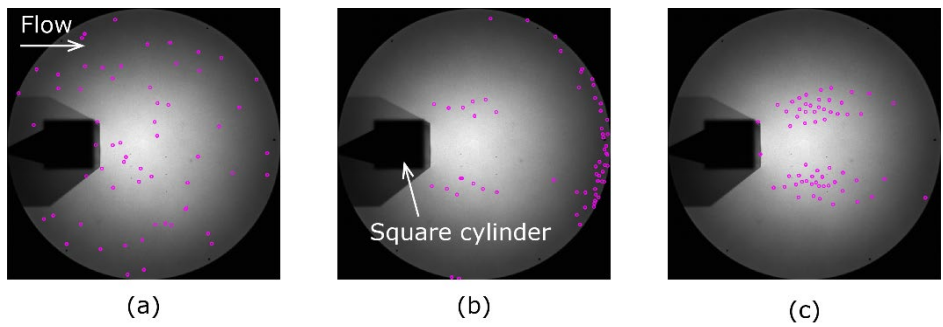

**Figure 6.** Typical results of the sensor points ($q = 60$) calculated by (a) random selection method, (b) the existing method and (c) the proposed method.

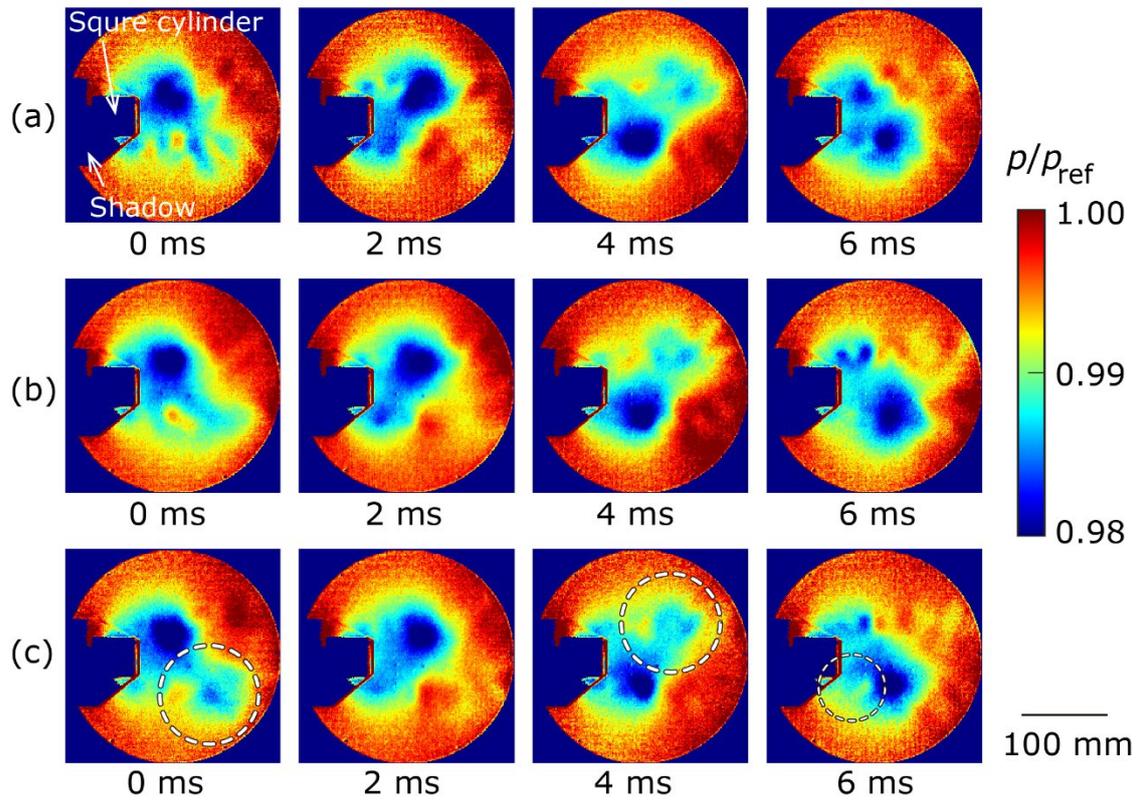

**Figure 7.** Typical reconstructed pressure distributions for the sensor points ($q = 60$) calculated by (a) random selection method, (b) the existing method, and (c) the proposed method. The flow direction is left to right.

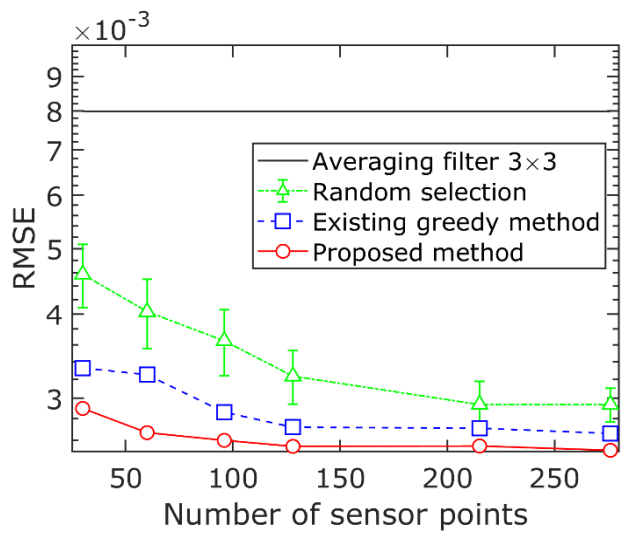

**Figure 8.** The comparison of RMSE between the proposed method and other methods

**Supporting Information**

Supporting Information is available from the ******* or from the author.

**Acknowledgements**

The authors wish to thank Dr. Yasuhumi Konishi, Mr. Hiroyuki Okuizumi, Mr. Hiroya Ogura, Mr. Benmao Lee, Mr. Kosuke Hayashi and Mr. Yuya Yamazaki during the wind tunnel testing at the Institute of Fluid Science, Tohoku University. We would like to thank Mr. Koyo Kubota for his assistance in the data processing. We also gratefully appreciate Tayca corp. for providing titanium dioxide. This work was partially supported by JST, PRESTO Grant Number JPMJPR187A, Japan. Part of the work was carried out under the Collaborative Research Project of the Institute of Fluid Science, Tohoku University.

**Conflict of Interest**

Y.N. and K.K. are employees of Fujitsu Ltd.

**Data Statement**

The data that support the findings of this study are available from the corresponding author upon reasonable request.

**Author Contributions**

**Tomoki Inoue** and **Yu Matsuda** contributed equally to this work.

**Tomoki Inoue**: Conceptualization, Data curation, Formal analysis, Investigation, Methodology, Software, Validation, Visualization, Writing-original draft.

**Tsubasa Ikami**: Data curation, Investigation, Resources, Visualization, Writing -review & editing

**Yasuhiro Egami**: Data curation, Formal analysis, Investigation, Resources, Writing -review & editing

**Hiroki Nagai**: Investigation, Resources, Writing -review & editing

**Yasuo Naganuma**: Methodology, Software, Validation

**Koichi Kimura**: Methodology, Software, Validation

**Yu Matsuda**: Conceptualization, Data curation, Formal analysis, Funding acquisition, Investigation, Methodology, Project administration, Resources, Software, Supervision, Validation, Visualization, Writing – original draft, Writing – review & editing

# Supporting Information

## Data-Driven Optimal Sensor Placement for High-Dimensional System Using Annealing Machine

*Tomoki Inoue* [1], *Tsubasa Ikami* [2,3], *Yasuhiro Egami* [4], *Hiroki Nagai* [3], *Yasuo Naganuma* [5], *Koichi Kimura* [5], *Yu Matsuda* [1,6,*]

1. Department of Modern Mechanical Engineering, Waseda University, 3-4-1 Ookubo, Shinjuku-ku, Tokyo, 169-8555, Japan

2. Department of Aerospace Engineering, Tohoku University, 6-6-01 Aoba, Arakaki, Aoba-ku, Sendai, Miyagi-Prefecture 980-8579, Japan

3. Institute of Fluid Science, Tohoku University, 2-1-1 Katahira, Aoba-ku, Sendai, Miyagi-prefecture 980-8577, Japan

4. Department of Mechanical Engineering, Aichi Institute of Technology, 1247 Yachigusa, Yakusa-Cho, Toyota, Aichi-prefecture 470-0392, Japan

5. Quantum Software Project, Quantum Laboratory, Research Unit of Fujitsu Research, Fujitsu Ltd. Kanagawa 211-8588, Japan

6. Japan Science and Technology Agency, PRESTO, Saitama 332-0012, Japan

* Corresponding author: y.matsuda@waseda.jp

**Figure S1** shows typical pressure distibutions (raw data) measured by the pressure-sensitive paint method, where the pressure $p$ is normalized by an atmospheric pressure $p_{\text{ref}}$. In our experimental condition, the frequency of the Kármán vortex was approximately 125 Hz, and we show a one periodic pressure distribution. **Figure S2** shows the calculated sensor positions. In the figure, the points calculated by the proposed method and those by random selection method and existing greedy method were compared. As discussed in the main text, the sensor points at the downstream of the squared cylinder were selected by the proposed method. **Figures S3 to S5** show the reconstructed pressure distributions based on the random selection, the existing greedy method, and the proposed methods. The noise in pressure distribution decreased with increasing the number of sensor points. For each method, a noticeable discrepancy is observed when the number of sensor points is small, i.e., the shape of the vortex (blue reagion) at 7 ms for the number of sensors is 30 is blurred in the data obtained by the random selection method and the existing greedy method. This vortex is clearly captured by the data obtained by the proposed method.

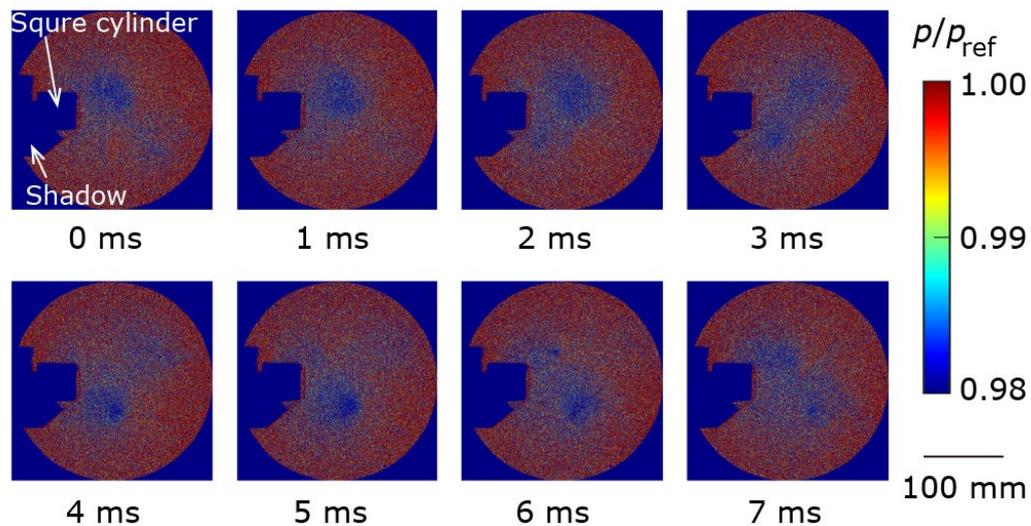

**Figure S1.** Typical pressure distributions (raw data). The pressure $p$ is normalized by an atmospheric pressure $p_{\text{ref}}$.

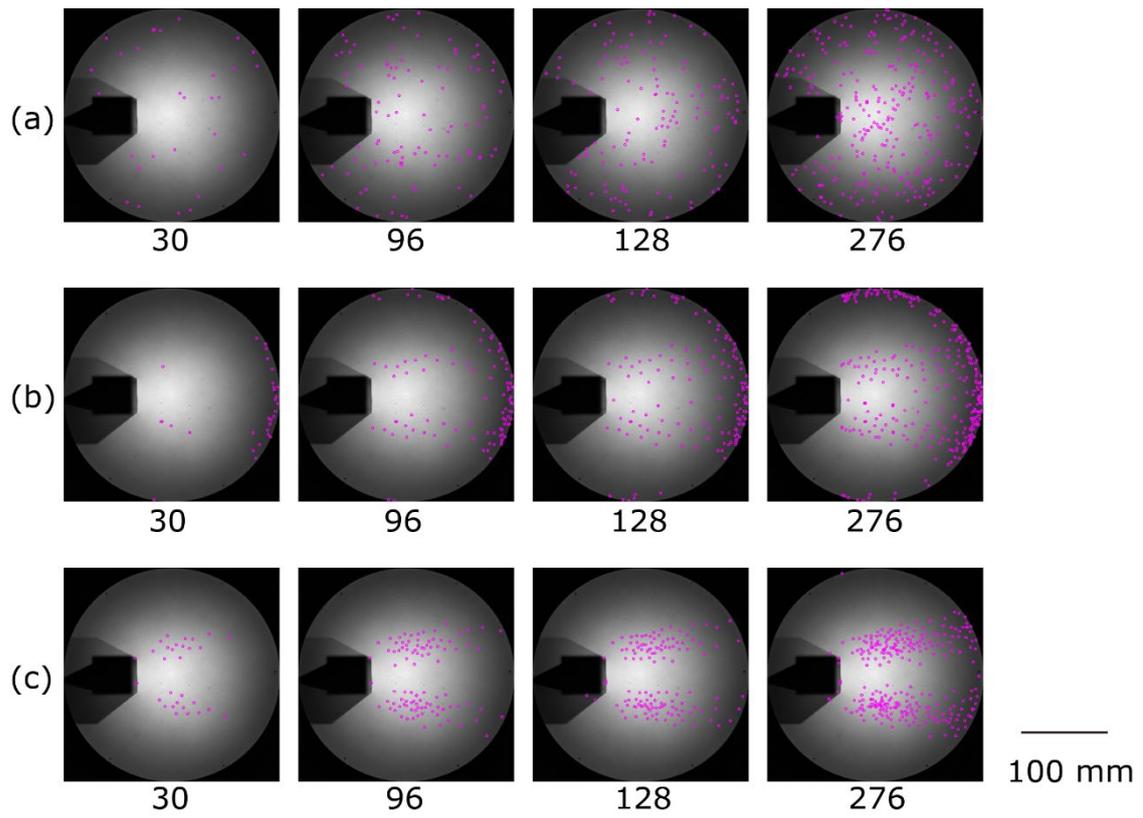

**Figure S2.** Typical results of the sensor points of 30, 96, 128, and 276 calculated by (a) the random selection method, (b) the existing method, and (c) the proposed method. The flow direction is left to right.

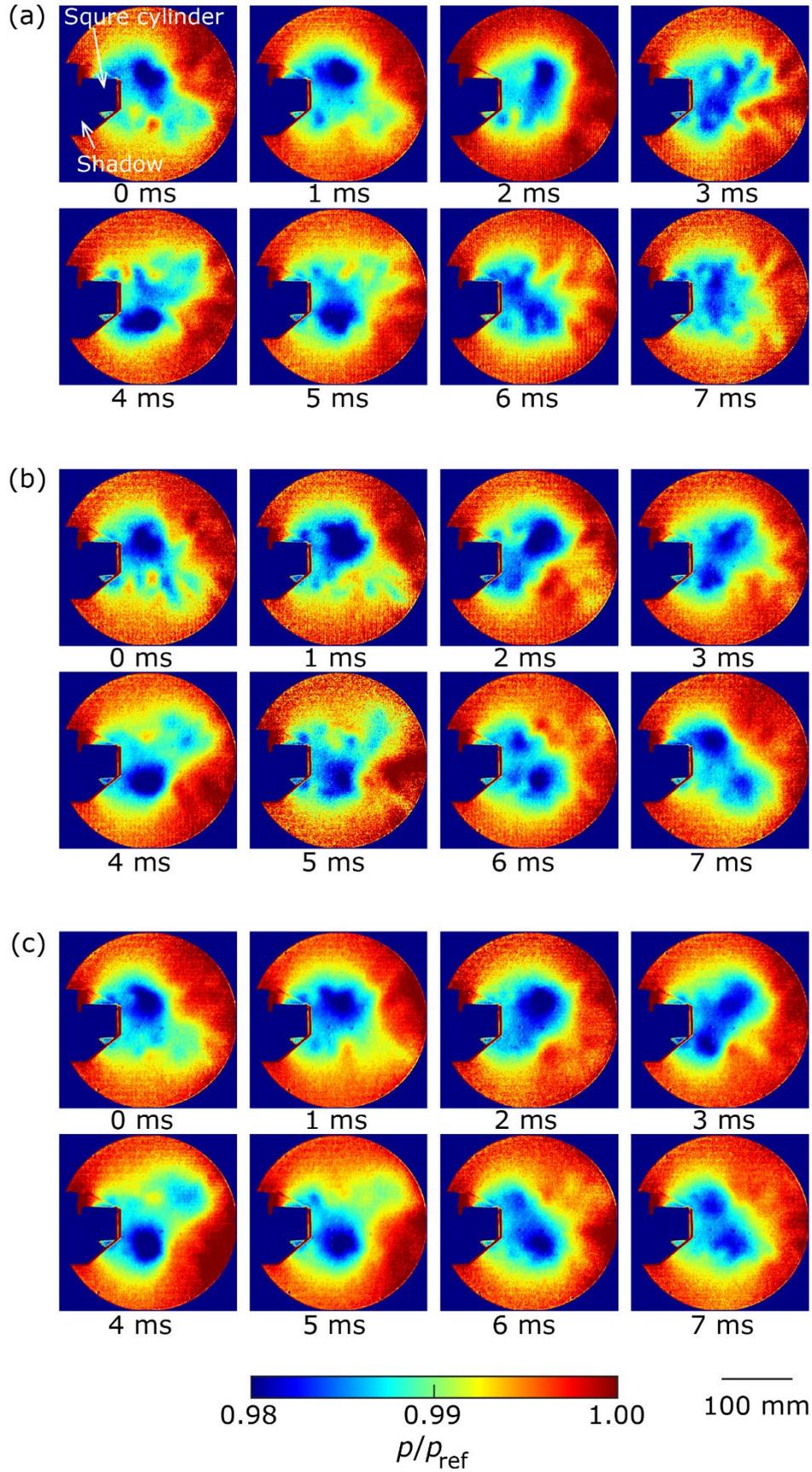

**Figure S3.** Reconstructed Pressure distribution based on the number of sensor points (a) 30, (b) 60, and (c) 128 calculated by the random selection method

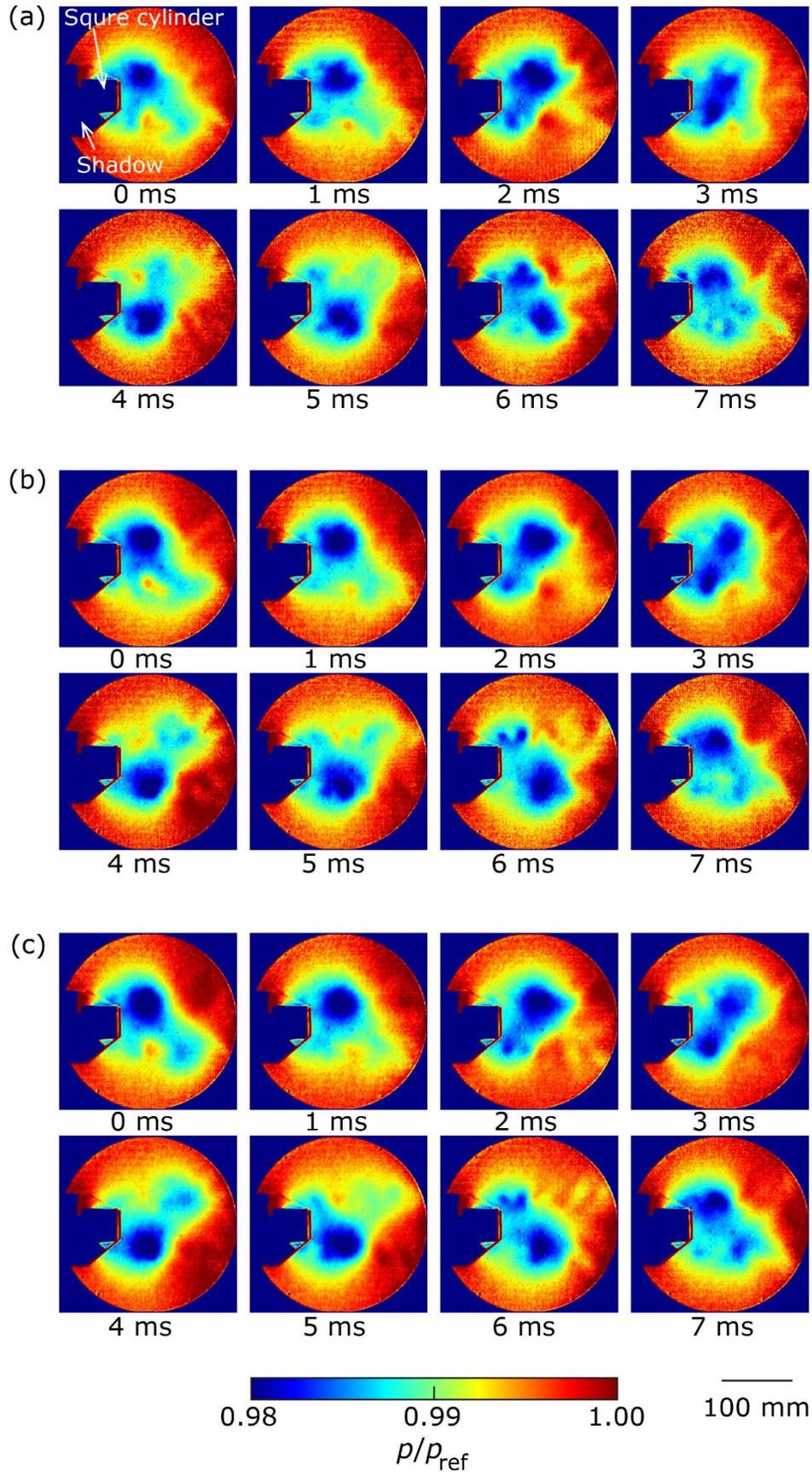

**Figure S4.** Reconstructed Pressure distribution based on the number of sensor points (a) 30, (b) 60, and (c) 128 calculated by the existing greedy method

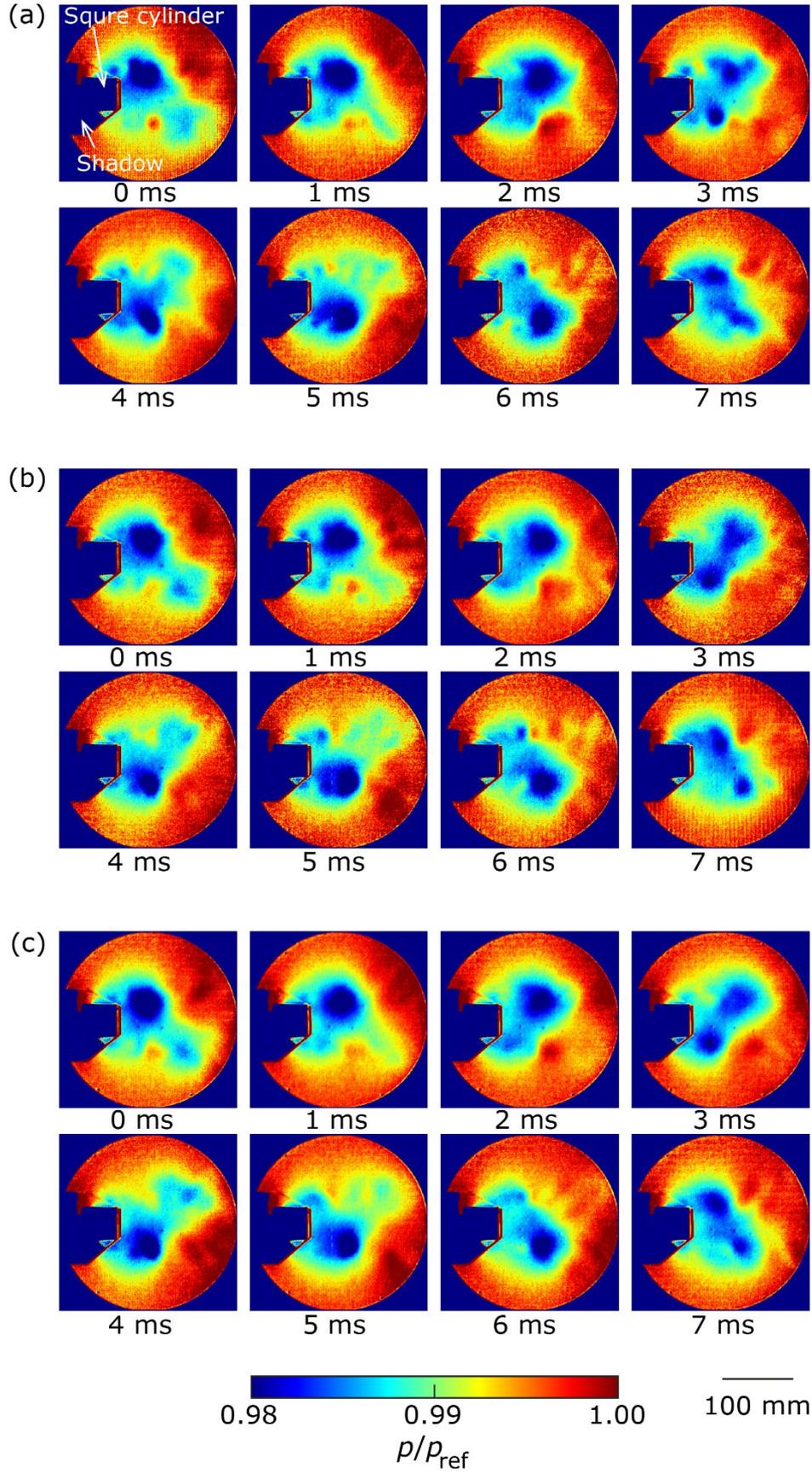

**Figure S5.** Reconstructed Pressure distribution based on the number of sensor points (a) 30, (b) 60, and (c) 128 calculated by the proposed method